\makeatletter
\declare@file@substitution{revtex4-1.cls}{revtex4-2.cls}
\makeatother

\documentclass[preprint]{aastex631}

\usepackage{natbib,twoopt}
\bibpunct{(}{)}{;}{a}{}{,} 
\usepackage{xcolor}
\usepackage[generate,ps2eps]{abc}
\renewcommand{\arraystretch}{1.5}
\usepackage{lmodern,babel,booktabs,multirow} 
\usepackage[T1]{fontenc}

\begin{document}

\title{Relative yield of thermal and nonthermal emission during weak flares observed by STIX during September 20--25, 2021}

\author[0000-0001-5313-1125]{Arun Kumar Awasthi}
\affiliation{Space Research Centre, Polish Academy of Sciences, Bartycka 18A, 00-716 Warsaw, Poland}

\author[0000-0003-4142-366X]{Tomasz Mrozek}
\affiliation{Space Research Centre, Polish Academy of Sciences, Bartycka 18A, 00-716 Warsaw, Poland}

\author{Sylwester Ko\l oma\'nski}
\affiliation{Astronomical Institute, University of Wroclaw, Kopernika 11, 51-622 Wroclaw, Poland}

\author{Michalina Litwicka}
\affiliation{University of Wrocław, Centre of Scientific Excellence – Solar and Stellar Activity, Joliot-Curie 12, PL-50383 Wroclaw, Poland}
\affiliation{Space Research Centre, Polish Academy of Sciences, Bartycka 18A, 00-716 Warsaw, Poland}

\author{Marek St\k{e}\'slicki}
\affiliation{Space Research Centre, Polish Academy of Sciences, Bartycka 18A, 00-716 Warsaw, Poland}

\author{Karol Ku\l aga}
\affiliation{Astronomical Institute, University of Wroclaw, Kopernika 11, 51-622 Wroclaw, Poland}

\begin{abstract}

 The disparate nature of thermal-nonthermal energy partition during flares, particularly during weak flares, is still an open issue. Following the Neupert effect, quantifying the relative yield of X-ray emission in different energy bands can enable inferring the underlying energy release mechanism. During September 20-25, 2021, the Solar Orbiter mission - being closer to the Sun ($\sim$0.6 AU) and having a moderate separation angle (<40$^{\circ}$) from the Sun-Earth line provided a unique opportunity to analyze multi-wavelength emission from $\sim$200 (mostly weak) flares, commonly observed by the Spectrometer Telescope for Imaging X-rays (STIX), STEREO-A, GOES, and SDO observatories. Associating the quotient (q$_{f}$) of hard X-ray fluence (12-20 keV) and soft X-ray flux (4-10 keV) with the peak SXR flux enabled us to identify strongly non-thermal flares. Multi-wavelength investigation of spectral and imaging mode observations of the 20 strongly non-thermal weak flares reveals an inverse relationship of q$_{f}$ with the emission measure (EM) (and density), and a positive relationship with the flare plasma temperature. This indicates that plasma in tenuous loops attains higher temperatures compared to that in the denser loops, in response to nonthermal energy deposition. This is in agreement with the plasma parameters of the coronal loops, as derived by applying the one-dimensional Palermo Harvard (PH) hydrodynamical code to the coronal loop plasma having different initial coronal loop base pressures when subjected to similar heating input. Our investigation, therefore, indicates that the plasma parameters of the flaring loop in the initial phase have a decisive role in thermal-nonthermal energy partitioning.
 \end{abstract}
	

\section{Introduction}
    Solar flares and accompanying events are one of the most energetic and dynamic phenomena in the Solar System emitting a substantial amount of energy in the X-ray and Extreme Ultraviolet (EUV) wavelength range. Being one of the first observational signatures of solar eruption, flares not only enable probing the plasma processes at play in the million-kelvin solar atmosphere but also improve our capability of predicting the magnitude and timing of the impact of solar eruptions on the Earth and beyond. Therefore, an in-depth analysis of multi-wavelength emission from the Sun can enable us to infer the plasma processes in the solar atmosphere and observationally unexplored characteristics of stellar flares and coronal mass ejection events \citep{Argiroffi2019, Veronig2021}.
    
    Flare emission covers the entire electromagnetic spectrum with the high-temperature (million--kelvin) plasma emitting in the X-ray and extreme ultra-violet (EUV) wavelengths while the low-temperature plasma is best seen in the H$\alpha$ and optical wavelengths. According to the standard model \citep[CSHKP:][]{Carmichael1964, Sturrock1966, Hirayama1974, Kopp1976} of energy released during solar flares \citep[unified 2D \& 3D models in][respectively]{Shibata2011, Aulanier2012}, the energy produced following the magnetic reconnection is predominately utilized in heating the ambient plasma and accelerating particles. Such a non-thermal pool of particles (generally electrons) deposits their energy in the denser layers of the solar atmosphere and produces hard X-rays (HXRs). Thermal distribution of particles emitting soft X-rays (SXRs) is subsequently generated from the chromospheric evaporation-driven plasma. Due to this causal relationship, flares often exhibit an `empirical' temporal relationship between SXR and HXR emissions, also termed the empirical Neupert effect \citep[ENE;][]{Neupert1968}. However, a detailed quantification of the plasma parameters in the context of the Neupert effect enables shedding light on the plasma processes occurring in various layers of the solar atmosphere. For example, by investigating the differential emission measure (DEM) of 80 flares, recorded by YOHKOH/SXT and BCS instruments, \citet{McTiernan1999} found that primarily the flares with high-temperature plasma ($\ge$16.5MK) exhibited the Neupert effect. 
    
    The integral form of the Neupert effect, which states that time-integrated HXR emission should mimic SXR emission, enables assessing the thermal-nonthermal energy partition during flares. However, \citet{Lee1995} and \citet{Veronig2002b} questioned this \textit{simple} form of relationship between the SXR and HXR emission by demonstrating a discrepancy between the slopes of the HXR fluence distributions and soft X-ray flux distribution. It has been further argued that HXR and SXR emissions may not be a comprehensive indicator of thermal and nonthermal energies, and it is the respective energy contents that should exhibit the Neupert effect. Further, the proportionality between HXR fluence and SXR flux should also additionally depend on the flare plasma parameters, as shown by \citet{McTiernan1999}. Therefore, \citet{Veronig2005} formulated a theoretical Neupert effect (TNE) which associates the beam power supply (inferred from HXR emission), and power required for observed SXR emission. However, for simple approximations of loop geometry along with the application of particle and energy transport schemes, the results of the TNE relationship were found to be similar to ENE. 
    
    Based on a critical overview of thermal, nonthermal, and bolometric energy content, \citet{Warmuth2020}, inferred the dependence of thermal-nonthermal energy partition on the flare strength. They indicated that weak flares exhibit a deficit of energetic electrons in contrast to the strong flare for which the injected nonthermal energy has been found sufficient to account for the thermal energy content. Several case studies of weak flares revealed the extraordinary nature of HXR emission during microflares, for example, some flares show SXR emission without distinctive enhancement in the HXR flux \citep{Awasthi2014, Awasthi2018b} while other cases exhibit non-thermal electrons flux with a spectral slope as hard as 6 \citep{Wright2017, Awasthi2018a}. Evidently, weak flares often exhibit a departure from the Neupert effect, as clearly seen in the correlation plot of GOES peak flux and HXR fluence (from Hard X-Ray Burst Spectrometer (HXRBS) onboard Solar Maximum Mission (SMM) observations) in \citet{Veronig2002a} where weaker events exhibit a larger scatter. Conventionally, solar flares that release a huge amount of X-ray emission (e.g. X-class flares) are naturally believed to be extreme events in the context of space-weather impact. However, an extremely broad range of variability within thermal-nonthermal emissions during flares (e.g. strongly nonthermal flares \citep{Awasthi2018a, Lysenko2018, Lysenko2023}, thermal flares \citep{Battaglia2009, Altyntsev2012, Awasthi2014} reveals a gap in our understanding of the standard flare energy release scheme. More recently, \citet{Fleishman2022} found the magnetic reconnection to efficiently accelerate almost all the ambient electrons leaving the region depleted of thermal electrons. Such episodes of energy release and associated energy release processes should have an imprint on the emission recorded in X-rays, extreme ultra-violet (EUV), and H$\alpha$ wavelengths.
    
    Flare plasma parameters are also believed to affect the yield of HXR and SXR emission, and therefore play a decisive role in the thermal-nonthermal energy partition. According to \citet{Lee1995}, the proportionality coefficient between the HXR fluence and SXR flux may depend on the magnetic field configuration and viewing angle and thus may vary from flare to flare. Early impulsive (cold) flares, investigated by \citet{Lysenko2023} and \citet{Lysenko2018}, have been found to originate from relatively dense, and shorter loops with a stronger magnetic field compared to the plasma charactersitics and geometrical properties of average flares investigated. \citet{Saqri2023arXiv} found conclusive evidence of microflares with strong nonthermal emission to originate from the loops which have one of their footpoints rooted in the umbral or penumbral region of sunspots. \citet{Motorina2020} investigated the thermal-nonthermal energy partitioning during a flare that originated from a two-loop system. Their investigation has revealed that although the nonthermal energy is partitioned in the loops in a comparable amount, one of the loops contained plasma with a higher temperature and lesser density compared to that in the other loop. Such disparate thermal response of the flare loops, even after being subjected to similar energy input from nonthermal electrons, is attributed to different initial plasma parameters (e.g. density) of the loops. On the contrary, \citet{Fleishman2021}, based on the investigation of a three-loop solar flare, found an uneven distribution of nonthermal energy in the loops and suggested the initial distribution of thermal plasma to guide the proportioning of thermal and nonthermal energies. It is evident that initial flare plasma parameters and their subsequent evolution characteristics may be the key to better understanding the thermal-nonthermal energy partition during flares. Therefore, in this work, we first conduct a statistical investigation of the yield of SXR and HXR emissions for all the flares that occurred during September 20-25, 2021, as presented in section~\ref{subsec:rel_productivity_def}. Next, an in-depth multi-wavelength investigation of 20 weak flares, that exhibit strong HXR emission, is presented. We then employ one-dimensional Palermo-Harvard (PH) hydrodynamical code to explore the response of the flare plasma for different initial atmospheric conditions and present the results in section~\ref{sec:PH_model}. The discussion and conclusions are offered in section~\ref{sec:disc_conc}. 
        
\section{Observations and Instruments}
\begin{figure}
		\centering
		\includegraphics[height=\textwidth, angle=90]{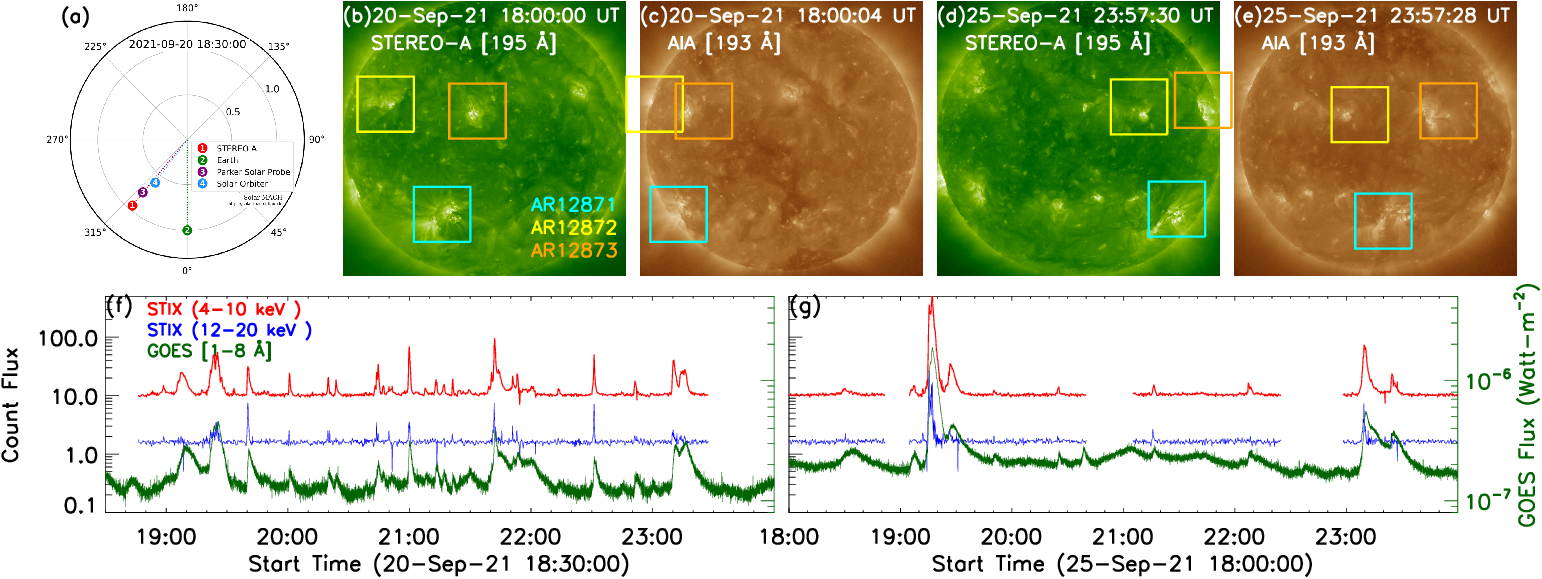}
		\caption{Overview of X-ray and EUV emission recorded by various instruments during September 20--25, 2021. (a) Positioning of various observatories on September 20, 2021. A sequence of images in 195{~\AA} from STEREO-A and 193{~\AA} from AIA on September 20, 2021, at 18:00 UT (b--c) and on September 25, 2021, at 23:57 UT (d--e) reveal overlap of field-of-view of STEREO-A and SDO. Active regions 12871, 12872, and 12873 have been marked. (e--f) X-ray intensity evolution in 4-10 keV, and 12-20 keV, from the quick-look mode observations from STIX instrument, and in 1--8{~\AA} from GOES mission, during 18:00--23:59 UT on September 20, 2021 (e), and during 18:00--23:59 UT on September 25, 2021 (f), demonstrating overlapping observations of flare events recorded by instruments positioned at different vantage points.}\label{fig:overview}
	\end{figure}

    During September 2021, observing lines-of-sight of Spectrometer Telescope for Imaging X-rays \citep[][STIX; at $\sim$0.6 AU]{Krucker2020} onboard Solar Orbiter mission \citep{Muller2020}, and STEREO-A \citep[][at $\sim$0.95 AU]{Kaiser2008} remained in an overlapping line-of-sight (Figure~\ref{fig:overview}a), thus offering a unique opportunity for a detailed multi-instrument investigation. Although STEREO-A and Solar-Orbiter have been in a perfect alignment on 18 September 2021, the STIX observations have become available only since $\sim$18:30 UT on 20 September 2021. Nevertheless, a very small separation angle between the \textit{Solar Orbiter} and \textit{STEREO-A} (between 2.5--9$^\circ$ during September 20--25, 2021) provided a unique opportunity to investigate the long-term evolution of the source active regions producing flares of different intensity class. Besides, a relatively moderate (<40$^\circ$) separation angle between the observing line-of-sight of STIX and the Sun-Earth line (Figure~\ref{fig:overview}a) allowed a very good coverage of STIX observed flares with the \emph{Geostationary Observational Environmental Satellite (GOES)} and \textit{Solar Dynamics Observatory} \citep[SDO;][]{Pesnell2012}. Therefore, we have conducted a multi-wavelength investigation of solar flares that occurred between September 20--25, 2021.
    
    Near-perfect alignment of Solar Orbiter and STEREO-A provided a co-temporal imaging perspective. Full-disk observations in 195{~\AA}, acquired by STEREO-A showed the presence of three active regions 12871, 12872, and 12873 on September 20, 2022 (figure~\ref{fig:overview}b), which remained visible from the STIX perspective until September 25, 2021 (figure~\ref{fig:overview}d). Further, a detailed investigation of EUV emission during flares is made by analyzing the observations from Atmospheric Imaging Assembly \citep[AIA;][]{Lemen2012} instrument onboard SDO. AIA provides the observations of full-disk Sun in seven EUV channels (94, 131, 171, 193, 211, 304, and 335{~\AA}) and two UV wavelengths (1600 and 1700{~\AA}) with a spatial and temporal cadence of 1.5 arcsecond and 12 s (24 s for UV channels), respectively. This enabled us to identify the source region of the flares observed by STIX and \textit{GOES} and derive the thermal characteristics of the flare plasma. The active regions observed by STEREO-A have also been observed by SDO for the analysis duration (figure~\ref{fig:overview}c\&e).

    STIX onboard the Solar Orbiter observes the Sun uninterruptedly in X-ray wavelength in a 4-150 keV energy range with a time cadence as high as 0.1 s. The incoming X-ray emission, when passed through 30 pairs of X-ray opaque Tungsten grids of varying pitch placed at different orientations before 30 pixelated (8 large pixels and 4 small pixels) detectors, forms a moir\'e pattern which can be further analyzed to identify the location, shape, and brightness of the source \citep{Giordano2014, Hayes2022, Massa2022}. STIX observations are provided in two formats, pixel data (containing observations from all the individual detectors and pixels), and spectrograms (pixel-integrated observations). Besides, observations are also available in the form of quick-look plots, background measurements, flare locations, calibration spectra, etc. We have made use of the spectrogram mode observations for the present investigation since these offer better coverage of flares. 
    
    A moderate separation angle of STIX with respect to the Sun-Earth line (between 30$^\circ$--38$^\circ$) allowed the investigation of flares observed by STIX from the \textit{GOES} as well. \textit{GOES} provides the disk-integrated X-ray intensity profile in 1--8{~\AA} and 0.5--4{~\AA} with a time cadence as high as 1 second. From the X-ray intensity profile in 4-10 keV, and 12-20 keV, from the quick-look mode observations made available from the STIX instrument, and in 1--8{~\AA} from GOES mission, on September 20, 2021, and September 25, 2021 (Figure~\ref{fig:overview}f--g), it is evident that a majority of flares have been commonly recorded by both the STIX and GOES observatories. The application of the solar flare identification technique \citep{Xiao2023} based on the STIX quick-look light curve in 4--10 keV, which also estimates the equivalent GOES intensity class, has revealed that a total of 217 flares occurred in the investigated duration, out of which only 13 flares are $\ge$ C-class. This indicates that the selected time duration contains predominately weak flares.
    
\section{Results}
\subsection{Relative yield of soft X-ray and hard X-ray emission during flares}\label{subsec:rel_productivity_def}
    According to the \textit{Empirical Neupert effect} (ENE), the time-derivative of the soft X-ray flux (F$_{SXR}$) often mimics the hard X-ray (F$_{HXR}$) emission profile. In the integral form, this relation can be expressed in the following form (\textit{c.f.} section 13.5.5 of \citet{Aschwanden2005},\citet{Veronig2005}).
    \begin{equation}\label{eq:TNE}
        F_{SXR}(t_p) \propto \mathcal{F}_{HXR}
    \end{equation}
    Here, $F_{SXR}(t_p)$ corresponds to the soft X-ray (SXR) flux at the flare peak time (t$_p$) whereas $\mathcal{F}_{HXR}$ denotes the HXR fluence, derived by integrating the hard X-ray (HXR) flux (F$_{HXR}$) during the start (t$_0$) and the peak time (t$_p$) of the flare, as expressed following.    
    \begin{equation}
        \mathcal{F}_{HXR} = \int_{t_0}^{t_p} F_{HXR}(t) dt
    \end{equation}
    This formulation of ENE allows quantifying the yield of HXR emission relative to the SXR emission during a flare, defined as the quotient (q$_f$) of hard X-ray fluence and F$_{SXR}$ at the flare peak (t$_p$) as presented in equation~\ref{eq:qf}.     
    \begin{equation}\label{eq:qf}
        q_f(t_p)=\frac{\mathcal{F}_{HXR}}{F_{SXR}(t_p)}
    \end{equation}
    
    We consider 4--10 keV and 12--20 keV energy bands to characterize the soft X-ray (SXR) and hard X-ray (HXR) emissions, respectively. Temporal characteristics of the flare X-ray intensity profile (t$_0$, t$_p$, and rise-time (t$_r$), etc.), including the peak and background fluxes in the SXR and HXR energy bands, have been derived using a semi-automated scheme. This scheme starts with visually identifying the time duration enveloping a flare event (t$_{fl}$), as well as the time duration for estimating background flux (t$_{b}$). Next, the background fluxes in both energy bands are estimated by averaging the respective fluxes during t$_{b}$, and their standard deviations ($\sigma$) are also calculated. Start time (t$_0$) in each energy band corresponds to the time instance when the flux values exceed the background fluxes by 1$\sigma$ for the first time since the start of t$_{fl}$. Visually selected time durations, covering the flare maximum in both the energy bands, enable deriving the peak time (t$_p$) when the fluxes reach their peak values. In such a way, we have characterized the temporal evolution of a total of 203 flares, identified during the analysis period (September 20--25, 2021). 

\begin{figure}[h]
        \centering
        \includegraphics[height=0.8\textwidth, angle=90]{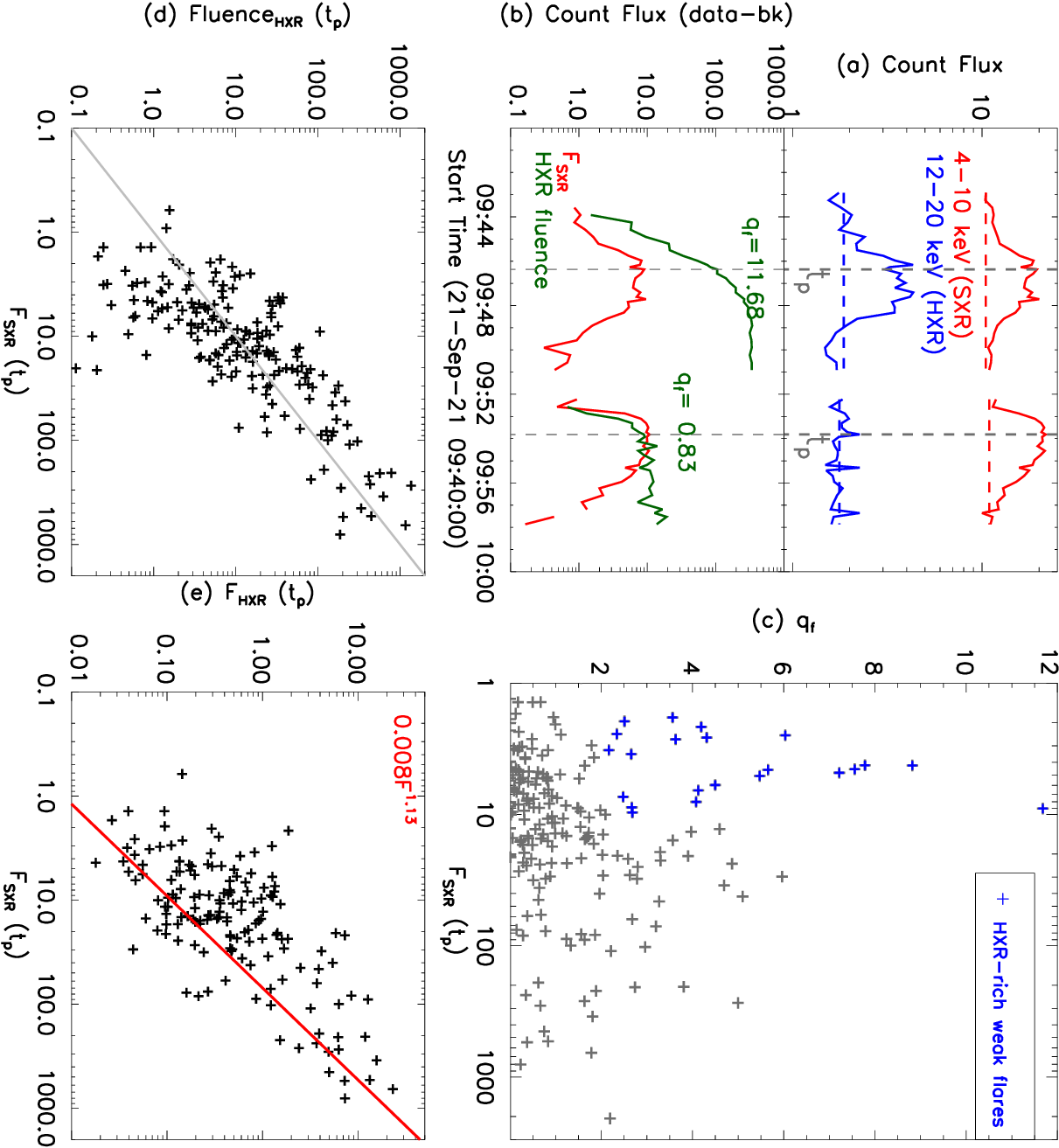}
        \caption{Statistical overview of hard X-ray flux yield (F$_{HXR}$; 12-20 keV) relative to soft X-ray Flux (F$_{SXR}$; 4-10 keV). (a) Two weak flares during 21-Sep-2021 09:40--10:00 UT in Soft and Hard X-ray energy bands, which exhibit the disproportionate nature of HXR emission irrespective of similar SXR emission. Estimated background levels and time of the F$_{SXR}$ maximum (t$_p$) have also been plotted. (b) Background-subtracted F$_{SXR}$, HXR-fluence, and quotient (q$_{f}$) of HXR-fluence and F$_{SXR}$ at t$_p$. (c) Variation of q$_{f}$ with respect to F$_{SXR}$(t$_p$). (d) F$_{SXR}$(t$_p$) versus F$_{HXR}$(t$_p$) exhibiting a power-law relationship.}\label{fig:qf_fsxr_hxr}
    \end{figure}

    Figure~\ref{fig:qf_fsxr_hxr}a shows the X-ray intensity profile of two weak flares that occurred on 21-Sep-2021 during 09:40--10:00 UT. Noticeably, despite similar peak soft X-ray flux values, hard X-ray emission during the flares is different. This disparate nature of relative X-ray emission in different energy bands can be quantified in terms of the quotient (q$_f$) (Figure~\ref{fig:qf_fsxr_hxr}b), as estimated to be 11.68 and 0.83, respectively for these flares. A higher value of q$_f$ indicates a relatively stronger hard X-ray emission. Such a disparate nature is indicative of a non-linear response of plasma environments toward the input nonthermal energy. In a similar fashion, q$_f$ has been determined for all the investigated flares by employing the background-subtracted flux in equation~\ref{eq:qf}. We find the q$_f$ values to be ranging between 0 (for 14 flares) and 11.68. According to the standard flare energy release model for long cooling times, q$_f$ should attain a value $\sim$1, which is indeed the case for 88 flares (corresponding to 0.5$\le$q$_f$$\le$2.5; 43\%). 
    
    We further categorize the relative yield of HXR emission with respect to the F$_{SXR}$(t$_p$) (Figure~\ref{fig:qf_fsxr_hxr}c). This representation enabled outlining that a significant fraction of the weak flares (F$_{SXR}$(t$_p$) <20 count/(s-cm$^2$-keV)) (54 flare cases) exhibit relatively large HXR emission for q$_f$>2. These flare cases are hereafter termed as the HXR-rich weak flares. On the other hand, peak soft X-ray flux F$_{SXR}$(t$_p$) exhibits a good correlation with hard X-ray fluence ($\mathcal{F}_{HXR}$) (Figure~\ref{fig:qf_fsxr_hxr}d), except for the weak flares as evident by a cloud of scattered points below the unity line (grey diagonal line for F$_{SXR}$(t$_p$)=$\mathcal{F}_{HXR}$) for F$_{SXR}$(t$_p$) <20 count/s-cm$^2$-keV. This indicates that a majority of weak flares have a tendency to exhibit predominant thermal emissions. In agreement, \citet{Veronig2002a} also found a large scatter for weak events in the correlation of HXR fluences (derived from HXRBS/SMM observations) and GOES peak fluxes. The correlation between F$_{SXR}$(t$_p$) and F$_{HXR}$(t$_p$) is found to be well explained by a power-law function with index 1.13 (Figure~\ref{fig:qf_fsxr_hxr}e). For more than 4000 RHESSI observed microflares, \citet{Hannah2008} found the correlation of thermal and nonthermal flux to follow a power-law relationship with an index of 1.11. 
    
\subsection{Temporal and spectral properties of plasma emission during HXR-rich weak flares} 
\begin{figure}[h]
        \centering
        \includegraphics[height=0.8\textwidth, angle=90]{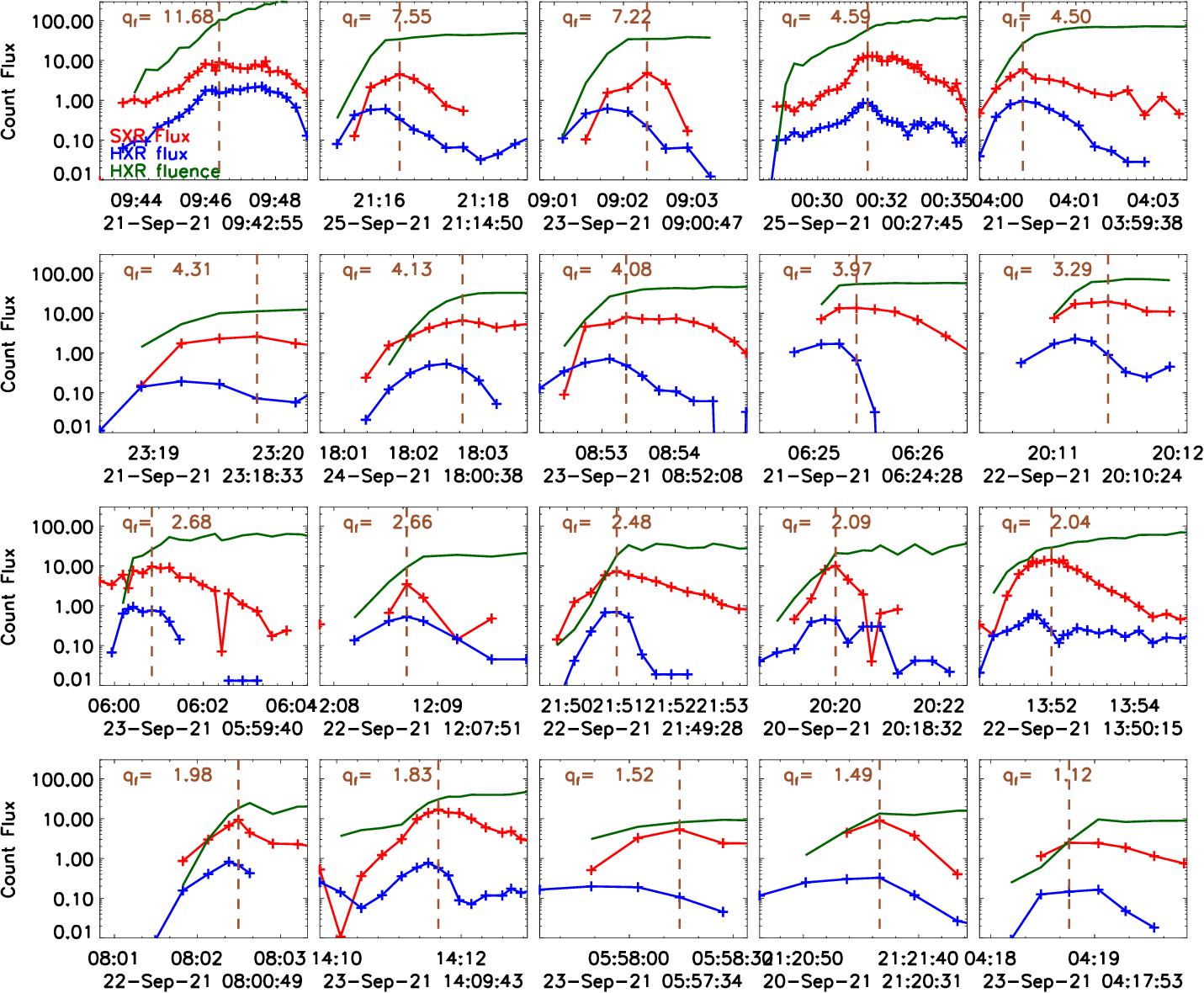}
        \caption{X-ray intensity profile of HXR-rich weak flares (q$_f$>1; F$_{SXR}$(t$_p$)$\sim$20 counts s$^{-1}$ cm$^{-2}$ keV$^{-1}$ ), selected for further multi-wavelength investigation and arranged in the descending order of q$_{f}$ value. F$_{SXR}$, F$_{HXR}$, and $\mathcal{F}_{HXR}$ are plotted in red, blue, and green, respectively. The time of F$_{SXR}$ peak emission (t$_p$), when q$_f$ is determined, is marked with vertical brown lines.}\label{fig:x-ray-lc-weak-flares-nth}
    \end{figure}

    The variation of the relative yield of HXR emission (q$_{f}$) with respect to the peak SXR flux enabled us to select 20 \textit{HXR-rich weak flare} cases (plotted in blue in figure~\ref{fig:qf_fsxr_hxr}c) for a further in-depth multi-wavelength investigation, as listed in Table~\ref{tab:nth_weak_table}. Flare intensity profiles in SXR (4-10 keV), and HXR (12-20 keV) energy bands, as determined from the STIX spectrograms, are plotted in figure~\ref{fig:x-ray-lc-weak-flares-nth}, and arranged in the descending order of q$_{f}$ value. 
    
    \begin{table}
    \setlength{\tabcolsep}{5pt}
    \caption{\label{tab:nth_weak_table}{Temporal characteristics of the SXR and HXR emission from HXR-rich weak flares and plasma properties during the SXR emission peak (t$_{p}$) as derived from the analysis of X-ray and EUV emissions}}
    \centering
    \begin{tabular}{cccccccc}
    \hline
    \multicolumn{1}{c}{\multirow{2}{*}{S.N.}} &
    \multirow{2}{*}{Flare} &
    \multirow{2}{*}{F$_{SXR}$\footnote{flux including pre-flare background, respective background values are provided in brackets}} &
    \multirow{2}{*}{q$_f$} &
    \multicolumn{2}{c}{EM (10$^{46}$cm$^{-3}$)} &
    \multicolumn{2}{c}{T (MK)} \\
    \cmidrule(lr){5-6} \cmidrule(lr){7-8}
     &  &  &  & X-ray & EUV\footnote{averaged over the flaring region} &  X-ray & EUV \\
    \hline
1 &  SOL2021-09-21T09:46:22 & 19.5(10.5) & 11.68 &  1.593 &  0.002 & 26.0 &  8.5 \\
2 &  SOL2021-09-25T21:16:23 & 15.3(10.8) &  7.55 &  0.919 &  0.026 & 11.8 & 11.1 \\
3 &  SOL2021-09-23T09:02:20 & 15.9(11.1) &  7.22 &  0.273 &  0.090 & 11.2 & 11.1 \\
4 &  SOL2021-09-25T00:31:55 & 23.3(10.4) &  4.59 &  0.859 &  0.033 & 10.9 &  8.5 \\
5 &  SOL2021-09-21T04:00:28 & 16.9(10.9) &  4.50 &  0.021 &  0.004 & 12.6 &  9.4 \\
6 &  SOL2021-09-21T23:19:49 & 13.4(10.8) &  4.31 &  1.103 &  0.009 & 10.5 &  8.5 \\
7 &  SOL2021-09-24T18:02:42 & 17.4(10.8) &  4.13 &  0.856 &  0.006 & 11.6 & 10.6 \\
8 &  SOL2021-09-23T08:53:19 & 18.7(10.7) &  4.08 &  0.035 &  0.015 & 11.5 & 11.0 \\
9 &  SOL2021-09-21T06:25:24 & 24.4(10.9) &  3.97 &  1.996 &  0.019 & 11.4 &  8.5 \\
10 &  SOL2021-09-22T20:11:26 & 30.9(11.5) &  3.29 &  1.867 &  0.227 & 12.8 & 11.2 \\
11 &  SOL2021-09-23T06:00:50 & 21.3(11.6) &  2.68 &  2.414 &  0.045 & 10.5 & 11.0 \\
12 &  SOL2021-09-22T12:08:42 & 14.0(10.5) &  2.66 &  9.479 &  0.022 & 16.2 & 10.6 \\
13 &  SOL2021-09-22T21:50:57 & 17.8(10.4) &  2.48 &  1.782 &  0.025 & 10.0 &  9.3 \\
14 &  SOL2021-09-20T20:19:59 & 20.7(10.6) &  2.09 &  3.711 &  0.006 & 10.3 &  8.9 \\
15 &  SOL2021-09-22T13:51:59 & 24.7(10.5) &  2.04 &  1.246 &  0.019 & 11.5 &  9.4 \\
16 &  SOL2021-09-22T08:02:29 & 20.4(11.1) &  1.98 &  0.616 &  0.019 & 11.7 & 11.1 \\
17 &  SOL2021-09-23T14:11:38 & 26.9(10.2) &  1.83 &  2.326 &  0.019 & 11.2 &  8.0 \\
18 &  SOL2021-09-23T05:58:14 & 21.5(16.2) &  1.52 &  1.097 &  0.117 & 11.2 & 11.2 \\
19 &  SOL2021-09-20T21:21:23 & 19.4(10.4) &  1.49 &  2.075 &  0.004 & 11.7 & 11.2 \\
20 &  SOL2021-09-23T04:18:45 & 16.1(13.6) &  1.12 &  0.615 &  0.004 &  8.4 &  8.4 \\
    \hline
    \end{tabular}
    \end{table}
    
\begin{figure}[h]
        \centering
        \includegraphics[height=0.95\textwidth, angle=90]{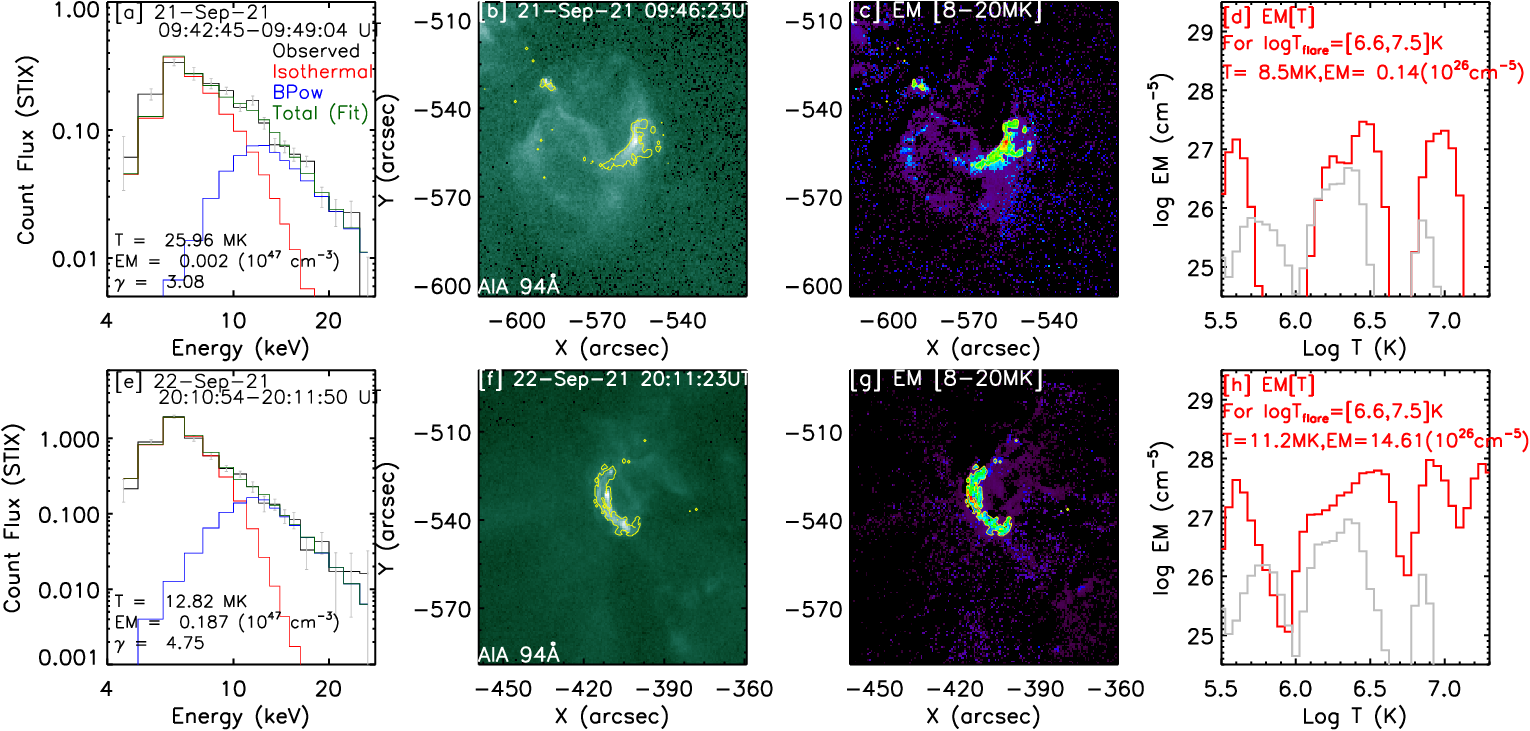}
        \caption{Thermal characteristics of the flare plasma as determined for HXR-rich weak flare case\#1 and \#10 from Table~\ref{tab:nth_weak_table}. (a) Forward fit of the observed X-ray spectrum (black) during the peak of SXR emission employing the isothermal (red) and broken power-law (blue) photon models. Flare plasma parameters that best fit the observations are annotated. 94~{\AA} image of the corresponding flare region (b), Emission measure map in 8--20 MK (c), and EM distribution (EM[T]) corresponding to the flaring region (average of EM[T] distribution over the pixels with EM > 10$^{26}$cm$^{-5}$ (yellow contours in (b) and (c)) in various temperature bins), and the non-flaring region (grey) is plotted in panel (d). Similarly, the bottom panels show the determination of flare plasma parameters from the X-ray (e) and EUV (f--h) emissions during flare case\#10. EM-weighted temperature, and average EM in the range logT=[6.6, 7.5]K for the respective flares are annotated in panels (d) and (h).  }\label{fig:xray_euv_thermal_nonthermal_analysis}
    \end{figure}

    Next, we investigate the thermal and nonthermal characteristics of the plasma during all the identified HXR-rich flares from the X-ray and EUV emission recorded by STIX and AIA instruments, respectively. Although the time durations of the flares range between 2 to 10 minutes, we have analyzed one representative X-ray spectrum per flare, which has been prepared with a time-integration ranging between 40--120 seconds covering the flare maximum phase. We fit the X-ray spectrum in 4--20 keV employing the isothermal and broken-power photon models available in the Object Spectral Executive (OSPEX) package of SolarSoftWare (SSW). OSPEX enables forward fitting of the observed X-ray spectrum by generating the theoretical spectrum from the selected photon model functions and convolving it with the instrument-specific spectral response matrix (SRM). The break energy in the broken power-law photon function is allowed to vary in the range of 10--15 keV, while the power-law index below the break energy is kept fixed to a very low value (equal to 1). This results in effectively providing fit results for a single power-law function with a cutoff at around 12 keV. In an iterative manner, the fitting aims to minimize the $\chi^2$ value, a measure of the difference between the observed and calculated count flux spectrum. We plot the spectral fit results for two investigated flare cases SOL2021-09-21T09:46:22UT and SOL2021-09-22T20:11:26UT (flare \# 1 \& 10 from table~\ref{tab:nth_weak_table}) in Figure~\ref{fig:xray_euv_thermal_nonthermal_analysis} (a \& e). 
    The best-fit spectral model enables deriving the thermal (temperature, emission measure) and non-thermal (photon spectral slope; $\gamma$) characteristics of the flare plasma. From the spectral fit, it is evident that the X-ray emission above 12 keV is best represented primarily with the nonthermal fit in our investigation. This supports our selection of energy bands that represent the thermal (SXR; 4--10 keV) and nonthermal (HXR: 12--20 keV) emission. Derived temperature and emission measure of the flare plasma for all the analyzed flares ranges between 8.4--26 MK, and 0.02--9.5 $\times$ 10$^{46}$cm$^{-3}$, respectively.

    The unavailability of pixel mode observations for a few flares restricted performing image synthesis from STIX observations. However, EUV images from AIA allow us to determine the morphological properties of the flaring region as well as the thermal characteristics of the flare plasma. Identification of the source region of weak flares is difficult, since several source regions of apparently similar brightness may be simultaneously present. To locate the source region of the flares, we first make use of the full-disk image of the Sun in 94~{\AA} at the time of maximum F$_{SXR}$ emission. For all the visually identified locations of enhanced brightening, we have then derived the EUV intensity profile from a sequence of 94~{\AA} cut-out images by making an average of the pixel intensities in all the pixels within the selected region that have an intensity larger than 20 DN s$^{-1}$. Subsequently, the temporal association between the X-ray and EUV intensity profiles has been made to distinctively identify the source region of the flares. From the 94~{\AA} images for flare\#1 and 10 from table~\ref{tab:nth_weak_table}~(Figure~\ref{fig:xray_euv_thermal_nonthermal_analysis} (b \& f)), recorded at the flare maximum, we note discrete kernel-like brightenings in addition to the extended loop-like emissions. Such discrete brightenings are believed to have resulted from the energy deposition from the nonthermal electron beams \citep{Awasthi2018b}. 
    
    To determine the thermal characteristics of the flare plasma from the EUV observations, we apply a modified sparse differential emission measure (DEM) inversion technique \citep{Cheung2015, SuY2018}. We employ this scheme on a sequence of EUV images in six wavelengths (94~{\AA}, 131~{\AA}, 171~{\AA}, 193~{\AA}, 211~{\AA}, and 335~{\AA}), acquired close-in-time around the SXR emission peak (t$_p$). This enables determining the emission measure distribution (EM[T]) in logT=[5.5, 7.5]K temperature range with binning of $\Delta$log$T$=0.05K. Synthesized EM maps corresponding to the 8-20 MK temperature bin reveal hot plasma of up to $\sim$20 MK temperature, co-spatial to the EUV brightening seen in 94~{\AA} images (Figure~\ref{fig:xray_euv_thermal_nonthermal_analysis} b--c \& f--g). This is in agreement with the flare plasma temperature as derived from the X-ray spectral fit. 
    
    From the 8--20 MK EM map, the region with EM values higher than 10$^{26}$cm$^{-5}$ is considered as the flare region (yellow contours in figure~\ref{fig:xray_euv_thermal_nonthermal_analysis}b--c \& f--g)). The area of the pixels that satisfy the aforesaid condition enabled determining the area (A) of the flare region. Further, assuming the flare region to be spherically symmetric and the estimated area (A) to represent its circular projection, its radius ((A/$\pi$)$^{1/2}$) may be regarded as the length equivalent (s) of the flare region \citep{Awasthi2016}. Subsequently, the volume of the flare plasma can be determined according to the relation, $V=\frac{4}{3}\pi s^3$.
    
    The EM[T] distribution of flare and quiet (EM < 10$^{26}$cm$^{-5}$) regions is determined by averaging respective values of EM[T] in all the temperature bins as plotted in red and grey, respectively, in figure~\ref{fig:xray_euv_thermal_nonthermal_analysis}d \& h. We note that the EM[T] distribution corresponding to the flare region exhibits a clear enhancement in the temperature range logT=[6.6, 7.5] in comparison to that corresponding to the quiet region. Therefore, we further estimate the EM-weighted temperature ($<T_{EM}>$) in the aforesaid temperature range (logT=[6.6, 7.5]) using the following relation.
    \begin{equation}
        T_{EM}=\frac{\sum_jT_j\times {\it {EM}}(T_j)}{\sum_j{EM}(T_j)}
    \end{equation}
    Here, EM[T$_j$] is the emission measure value corresponding to temperature T$_j$, and the averaging is performed over the temperature range logT=[6.6, 7.5]. A representative value of emission measure for each flare case (EM\_EUV) has been determined by averaging the EM[T] distribution over the aforementioned temperature range, and subsequently multiplying with the respective area of the flare region (A). We have provided the values of T$_{EM}$ and EM\_EUV in table~\ref{tab:nth_weak_table}. Temperature and EM values of the flare plasma from the EUV emission for all the investigated weak flare cases, determined as explained above, vary in the range of 8--11.2 MK, and 0.002--0.227 $\times$ 10$^{46}$cm$^{-3}$, respectively (see table~\ref{tab:nth_weak_table}). 

\begin{figure}[h]
        \centering
        \includegraphics[height=0.95\textwidth, angle=90]{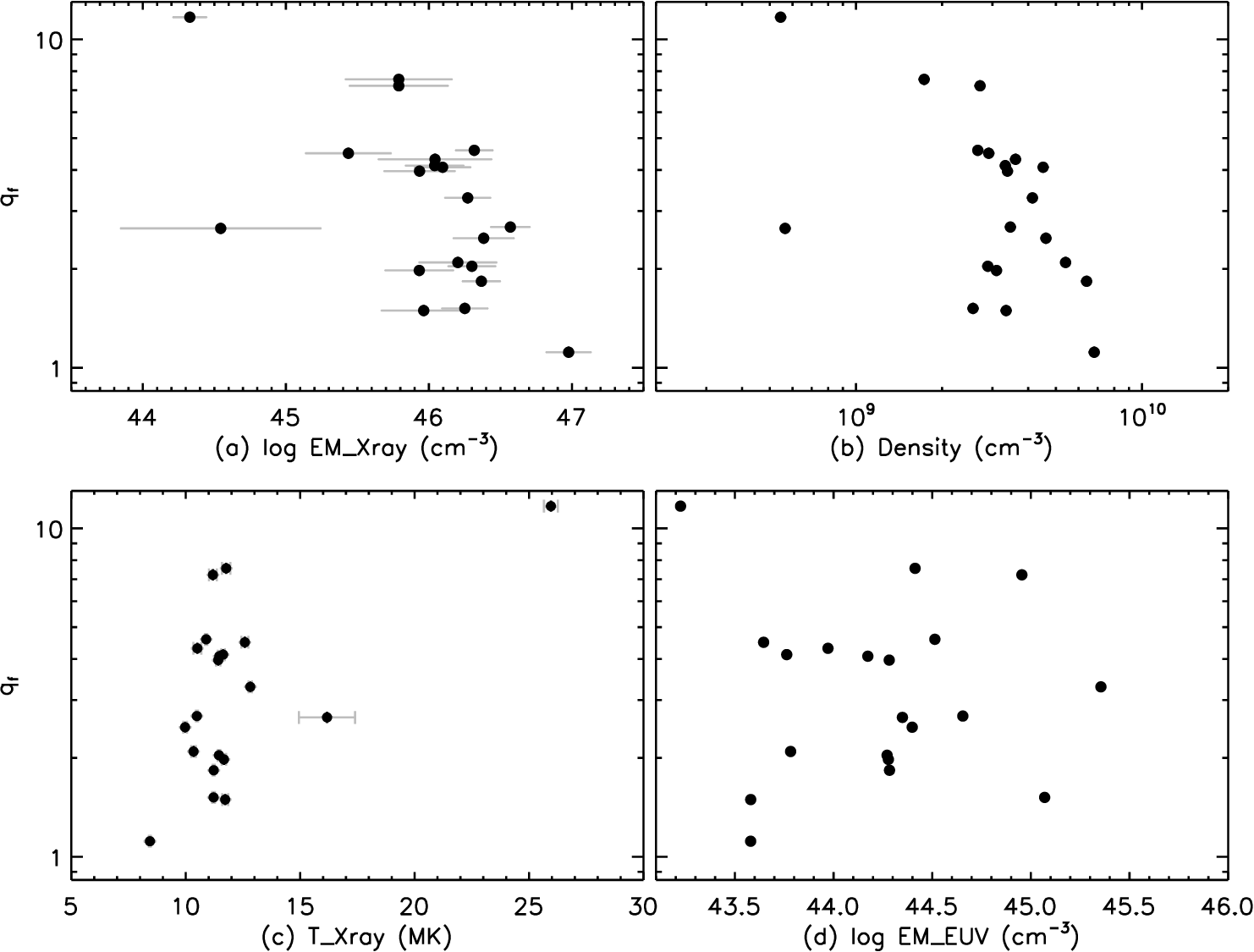}
        \caption{Correlation of q$_f$ with flare plasma parameters, such as emission measure (a), density (b),  temperature from analyzing X-ray emission (c), and EM derived from the EUV emission (d). The volumes estimated from the EUV images and EM$_{Xray}$ have been used for the density determination.}\label{fig:qf_t_em_ne}
    \end{figure}

    To determine the role of flare plasma parameters in the relative yield of HXR emission (q$_f$) in the weak flares, we have determined the correlation of temperature (T), emission measure (EM), and density (n$_e$) as determined from the X-ray as well as EUV observations (figure~\ref{fig:qf_t_em_ne}). The density of the flare plasma is estimated from the following relationship. 
    
    \begin{equation}
        n_e=\sqrt{\frac{EM_{Xray} (cm^{-3})}{V_{EUV} (cm^3)}}
    \end{equation}
    Here, $V_{EUV}$ is the volume of the flare region, estimated as discussed above. Since the image synthesis from X-ray observations has not been possible owing to the lack of `pixel mode' of observations for several flare cases investigated in this work, volumes estimated from the EUV images and EM$_{Xray}$ are used for the determination of respective density values. 
    
    We find the derived EM and n$_e$ to be inversely related with q$_f$, i.e., the emission measure of the flare plasma is lower in the case of flares exhibiting relatively larger HXR yield and vice-versa (figure~\ref{fig:qf_t_em_ne}a). This trend is evident in the density versus q$_f$ correlation plot as well (figure~\ref{fig:qf_t_em_ne}b). Further, plasma temperature is found to be positively correlated with the HXR-richness of the flares (figure~\ref{fig:qf_t_em_ne}c). These correlations indicate that the stronger nonthermal emission heats the flare plasma more efficiently without significantly increasing the plasma density. This implication is in agreement with the case studies performed by \citet{Motorina2020} and \citet{Fleishman2021}.

    Further, while we have derived the EM[T] distribution from the EUV emission corresponding to the flare plasma having a temperature larger than 4 MK (logT > 6.6 K), the EM (and temperature) determined from the EUV observations do not exhibit any clear correlation with q$_f$ (figure~\ref{fig:qf_t_em_ne}d).
    
    \subsection{Hydrodynamical simulation of flare plasma}\label{sec:PH_model}
    Although the analysis of X-ray and EUV observations in the present investigation indicate that flare plasma parameters systematically affect the partitioning of thermal and nonthermal emissions, it is not possible to determine the effect of initial plasma conditions on the energy partitioning, as indicated by \citet{Motorina2020} and \citet{Fleishman2021}, primarily due to low count statistics.  Hydrodynamical models, on the other hand, allow probing the response of the coronal loop plasma for different sets of initial conditions. Therefore, we perform the hydrodynamical simulations using the one-dimensional Palermo-Harvard (P-H) code \citep{Peres1982, Betta1997}. The PH code implements the strategy of solving the conservation equations of the mass, momentum, and energy numerically for the case of a compressible fluid. Further, the adoption of the adaptive grids enabled resolving complex conditions like extremely large temperature gradients in the transition region. A typical PH model run requires inputs such as loop length and a heating function varying along the loop and in time. This enables deducing the spatial and temporal evolution of temperature and density in the loop. 

    To determine the impact of varying plasma conditions on the response of coronal loop plasma that has been subjected to heating, we perform the PH simulation for 25 sets of models comprised of five different initial base pressures in the loop (10, 50, 100, 150, and 200 dyne cm$^{-2}$) and all are subjected to five values of heating inputs (5, 10, 15, 20, and 25 erg s$^{-1}$cm$^{-3}$). Other input parameters for the simulations have been the loop semi-length (L) = 10$^9$ cm, and the flare heating (E$_h$; spatial profile of the heating - Gaussian centered at 2 $\times$ 10$^8$ cm above foot point, and width (FWHM) L/5) has been applied during 100-130 sec of the simulation run, with a heating decay time = 10 s. The first step of the PH run requires creating a hydrostatic stable state loop plasma condition by subjecting the loop to a background heating (E$_{h0}$), which varies in the range of 0.0485 -- 1.16 erg s$^{-1}$cm$^{-3}$, depending on the desired initial base pressure. PH model is allowed to run for 1500 seconds. We have also tabulated the input parameters and heating function in Table~\ref{tab:ph_model}. 
    
    \begin{table}
    \caption{\label{tab:ph_model}Input parameters for PH model}
    \centering
    \begin{tabular}{ll}
    \hline
    Loop semi-length (L) & 1 $\times$ 10$^9$  cm  \\
    Non-flare background heating (E$_{h0}$) &  0.0485 -- 1.16 erg s$^{-1}$cm$^{-3}$ \\
    Flare heating (E$_h$) & [5, 10, 15, 20, 25] erg s$^{-1}$cm$^{-3}$   \\
    Spatial Profile of E$_h$ &  Gaussian centered at 2 $\times$ 10$^8$ cm above foot point, \\
                             &  and width (FWHM) L/5 \\
    heating decay time ($\tau$) & 10 s \\
    t$_{on}$--t$_{off}$ for E$_h$ & 100-130 s \\
    Base Pressure & [10, 50, 100, 150, 200] dyne cm$^{-2}$ \\
    \hline
    \end{tabular}
    \end{table}
        
    \begin{figure}[h]
        \centering
        \includegraphics[height=0.95\textwidth, angle=90]{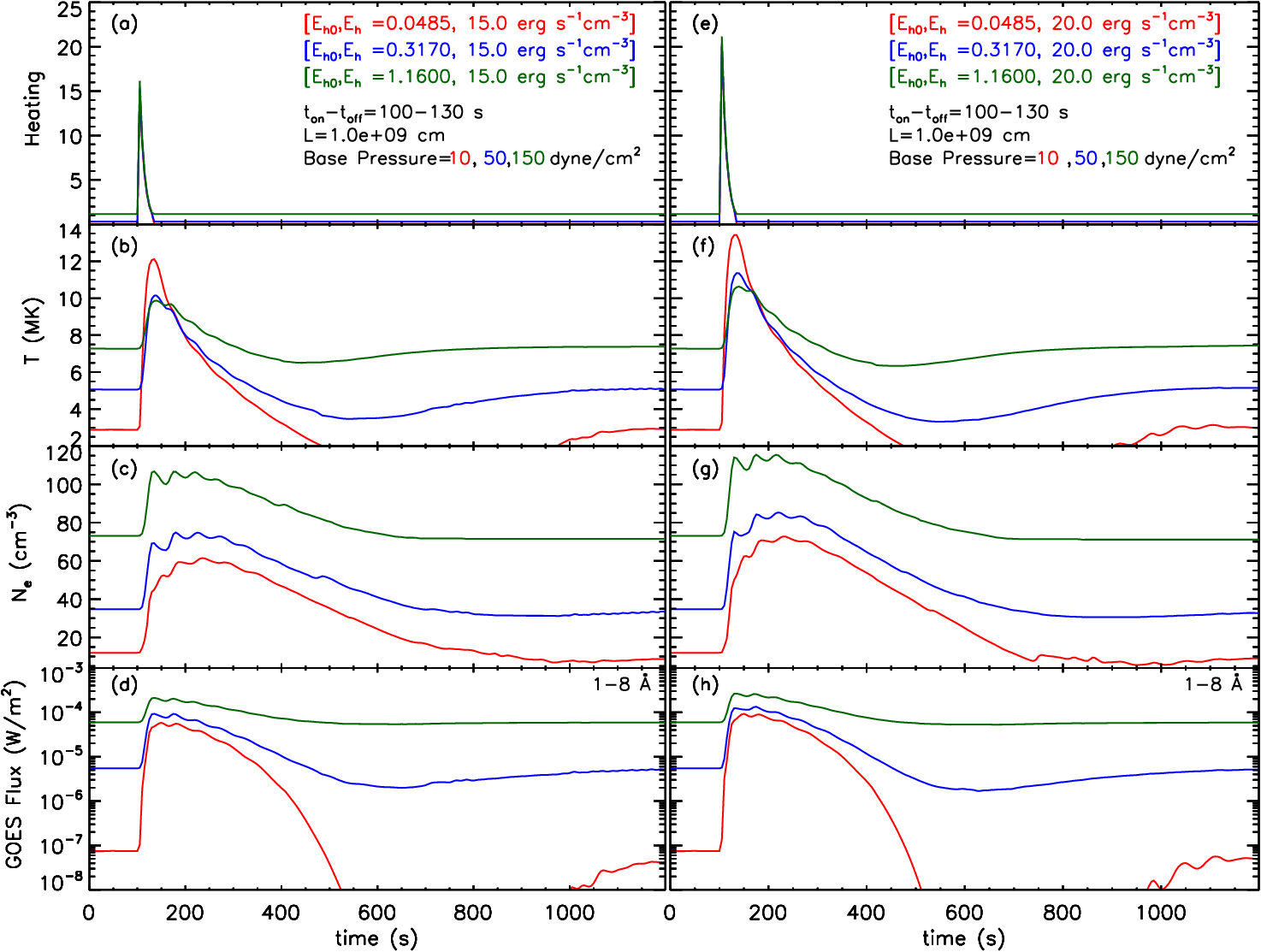}
        \caption{Evolution of plasma parameters at the loop apex as resulting from the application of the Palermo-Harvard hydrodynamical code. For two different heating inputs (15 and 20 erg s$^{-1}$ cm$^{-3}$) that are applied for 30 seconds starting at 100 seconds into the model run. The resulting temporal evolution profiles of temperature, density, and simulated GOES flux in 1--8{~\AA} wavelength band are shown in panels (a)--(d), \& panels (e)--(h), respectively.}\label{fig:PH_model_results}
    \end{figure}

    To assess the response of the loop plasma with different initial conditions when subjected to the same heating input, we plot the resulting plasma parameters at the loop apex for three different cases of initial base pressure when subjected to heating input of 15 and 20 erg s$^{-1}$ cm$^{-2}$ in the left and right panels of figure~\ref{fig:PH_model_results}, respectively. This analysis revealed that despite applying the same heating input to the loops with the same geometrical parameters, the model run for the case of lower base pressure resulted in a higher peak temperature (and lower density) compared to that of a higher base pressure. Similarly, the PH model run corresponding to the heating input of 20 erg s$^{-1}$ cm$^{-3}$ also revealed disparate temperature and density in the loops with different initial base pressure, even though subjected to the same heating input. Assuming the width of the loop as 5 arcseconds (w=2r; radius (r)), we have derived the emission measure (EM) using the plasma density (n$_e$) that has been obtained from the hydrodynamical simulation using the relationship EM = n$_e$$^2$V, where V is the volume and approximated to be equal to $\pi$r$^2$L considering the cylindrical geometry of the loop of length `L' and radius `r'. From the GOES fluxes in the long (1--8~{\AA}) and short (0.5--4~{\AA}) wavelength ranges, estimated using the EM and PH model-derived temperature, we find that the resulting soft X-ray flux in 1--8~{\AA} also strongly depends on the initial base pressure of the loop. This analysis asserts the role of initial loop plasma parameters in the observed X-ray emission during flares and hence in the thermal-nonthermal energy partitioning.

    \section{Discussion and Conclusions}\label{sec:disc_conc}
    While it is well known that non-potential energy available in the magnetic system is released during solar flares governed by the magnetic reconnection process, its partitioning between the thermal and nonthermal energies is not well understood. This is evident from  -- 1) dissimilar SXR and HXR emissions during a flare (e.g. figure~\ref{fig:qf_fsxr_hxr}a), and 2) observed disproportionate thermal and nonthermal emissions within the flare loops that originate from the same flare source region \citep{Motorina2020, Fleishman2021}. In addition to exposing our lack of knowledge of the energy release mechanism during flares, this poses questions if the flare intensity class, which is derived from the peak flux recorded by GOES in 1--8~{\AA} wavelength range, is a comprehensive quantifier of energy released during the flare. In this regard, this work characterizes the temporal evolution of SXR and HXR emissions during more than 200 flares, recorded by STIX during September 20-25, 2021. Using an integral formulation of the Neupert effect, we have quantified the relationship between SXR and HXR emission in the form of a quotient factor (q$_f$). q$_f$ represents the HXR fluence relative to the observed SXR emission at the flare maximum. This quantification enables identifying the HXR-rich (strongly nonthermal) flares from a large set of flares. By performing an in-depth multi-wavelength investigation of 20 HXR-rich weak flare cases, our analysis revealed that the relative HXR yield (q$_f$) is inversely correlated with the flare plasma density while exhibiting a positive correlation with the temperature. This indicates that in comparison to denser loops, flare loops containing plasma of lower density result in exhibiting relatively higher temperatures in response to similar nonthermal energy input. We further explore the response of heating the flare loop with different initial plasma conditions by the application of a one-dimensional Palermo-Harvard code. We have performed hydrodynamical simulations for various cases comprising five different initial base pressures in the coronal loops when subjected to five different values of heating inputs. This distinctively revealed that, for the same heating input, the loops with lower base pressure result in exhibiting plasma of higher temperature and lower density. This is in agreement with the implications made from the X-ray spectral analysis of HXR-rich flare cases.

    \citet{Motorina2020} conducted an in-depth analysis of the thermal response of a two-loop solar flare. They found that, despite the deposition of a comparable amount of nonthermal energy in both loops, the tenuous loop contained hotter plasma than the denser loop. Only one of a complex three-loop solar flare, investigated in \citet{Fleishman2021}, contained nonthermal electrons, while the other two loops contained only thermal plasma. More recently, \citet{Fleishman2022} found the magnetic reconnection to efficiently accelerate almost all the ambient electrons leaving the region to be depleted of thermal electrons. These investigations emphasize the importance of loop plasma conditions in thermal-nonthermal energy partitioning, which is in agreement with the inferences derived from our analysis. This further suggests that, in addition to investigating the temporal association of the SXR and HXR emission made in accordance with the Neupert effect, the density and temperature of the loop plasma should be quantitatively included for a comprehensive understanding of the energy release mechanism. 

\begin{figure}[h]
        \centering
        \includegraphics[height=0.8\textwidth, angle=90]{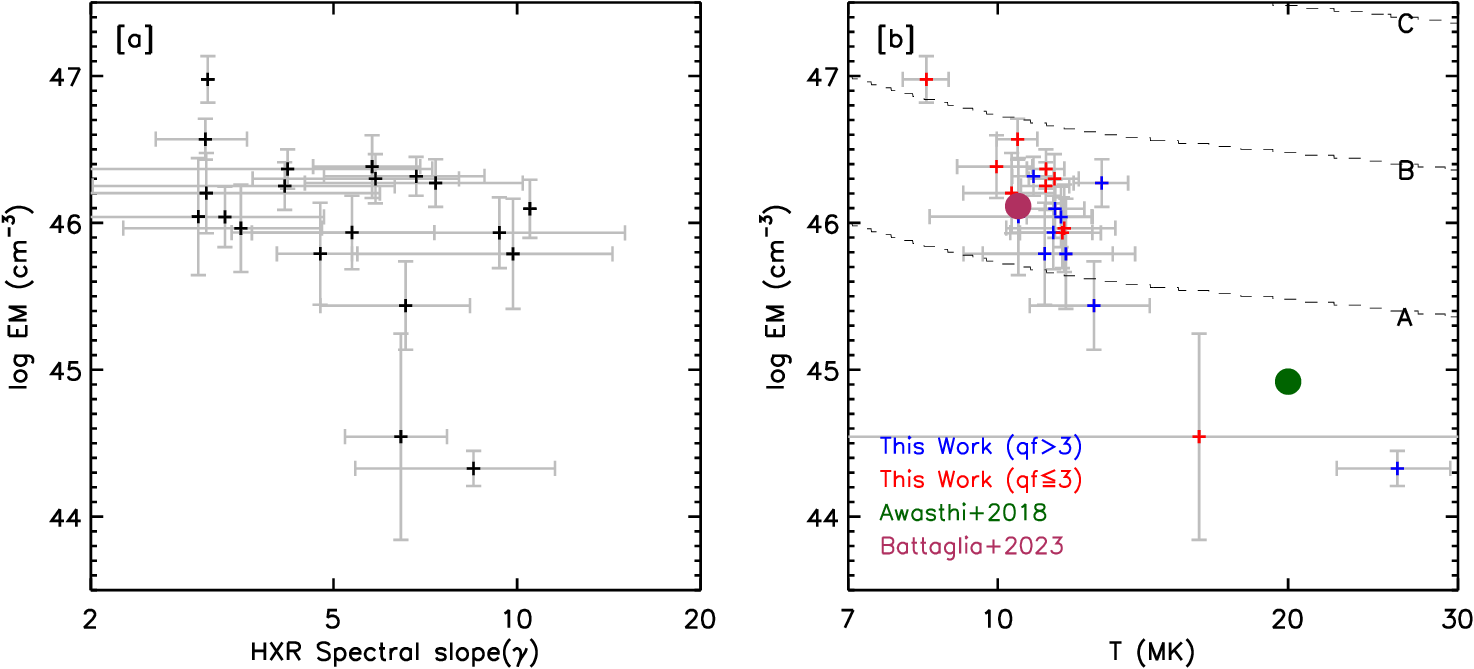}
        \caption{Variation of emission measure with HXR spectral slope (a) and temperature (b) for the HXR-rich weak flare cases. Flare cases with q$_f$>3 have been shown in blue, while the rest are plotted in red. ISO-flux lines for GOES A, B, and C classes are also drawn in grey. Relatively strong HXR-rich flares (blue) exhibit a trend of having a lower EM and a higher temperature. Flare plasma parameters for two strongly nonthermal weak flare cases analyzed in \citet{Awasthi2018a} and \citet{Battaglia2023} are also plotted.}\label{fig:goes_flux_t_versus_EM}
    \end{figure}

    In addition to the thermal characteristics of the flare plasma, we also derived the spectral slope ($\gamma$) of the HXR emission by forward fitting the observed X-ray spectra with combined isothermal and broken power-law photon models. We find the slope of nonthermal emission to inversely correlate with the emission measure (figure~\ref{fig:goes_flux_t_versus_EM}a). This implies that the nonthermal electrons with a harder spectrum result in an efficient chromospheric evaporation, as evidenced by higher EM values. While this result is in consensus with that revealed in the investigation of \citet{Motorina2020}, it is in disagreement with the simulation results of the chromospheric evaporation in response to the beam heating \citep{Fisher1985, Reep2015}, in which nonthermal electron with a softer spectrum results in an efficient evaporation. Despite this fact, the simulation results of \citet{Reep2015} also indicated that, for weak flares, an explosive evaporation threshold can be achieved with very little total nonthermal energy, and the thermal response of the atmosphere may not strongly depend on the electron energy in this regime. This may provide an explanation for a positive correlation of spectral hardness with the emission measure, obtained in the present work. 
    
    For the analyzed flares, we further deduce the correlation of temperature versus emission measure, as shown in figure ~\ref{fig:goes_flux_t_versus_EM}b. We may infer that relatively strong HXR-rich flares (blue) exhibit a trend of having a lower EM and a higher temperature. Flare plasma parameters corresponding to two strongly nonthermal weak flare cases (nonthermal electron spectral index $\delta$ = 4--6) analyzed in \citet{Awasthi2018a} and \citet{Battaglia2023} are overplotted which evidently follow the general behavior of T versus EM, obtained in the present work. Despite the fact that our work considers flares with similar peak SXR emission (< 20 counts s$^{-1}$cm$^{-2}$keV$^{-1}$), from the ISO-flux lines for GOES A, B, and C classes, drawn in figure~\ref{fig:goes_flux_t_versus_EM}b, the intensity classes of the analyzed flares is found to vary from sub-A to B-class. This representation further reveals the strong dependence of flare intensity class on the thermal-nonthermal energy partition. 
       
    A comparative overview of temperature values derived from the X-ray (T$_{X-ray}$) and EUV (T$_{EUV}$) observations reveals a significant difference for a few cases while results in similar values for others (Table~\ref{tab:nth_weak_table}). This behavior may be attributed to the non-equilibrium ionization (NEI). For heat pulses of less than a minute duration, \citet{Reale2008} found that NEI may result in much lower plasma temperature as determined from the observed spectra despite the source electrons of more than 10 MK temperature. \citet{Lee2019} investigated the effect of NEI on the thermal characteristics of the plasma, which was initially in the state of ionization equilibrium (EI), and subjected to rapid heating due to shock or magnetic reconnection. They found that the temperature of the plasma in NEI is underestimated, compared to that in EI. Therefore, a short heating of a coronal loop containing plasma of lower initial density and temperature may lead to different responses as seen in the AIA's EUV channels owing to the fact that the highly ionized atoms, which are needed to produce emission lines seen by AIA, will be in lower abundances. Further, the time needed to attain a higher degree of ionization may be as long as a few hundred seconds. 
        
    Summarily, this work emphasizes the crucial role of plasma density in the coronal loops in the context of the thermal-nonthermal energy partition. While it is difficult to determine initial plasma conditions solely from the X-ray spectral observations due to the limited signal-to-noise ratio in the rising phase of the flare, conducting case studies through multi-wavelength images will help unravel the exact nature of initial plasma conditions on the thermal response of the flare plasma and energy partition. Further, characterizing the onset temperature (e.g., \citet{daSilva2023}) and emission measure for flares of different intensity in a statistical sense may enable assessing their quantitative impact on thermal-nonthermal energy partition of the flare and is planned to be conducted in the future.   
    
\begin{acknowledgements}
    This research is supported by the PASIFIC project that received funding from the European Union's Horizon 2020 research and innovation programme under the Maria Sk\l odowska-Curie grant agreement No. 847639. This research was funded in part by the National Science Centre, Poland grant No. 2020/39/B/ST9/01591. For the purpose of Open Access, the author has applied a CC-BY public copyright licence to any Author Accepted Manuscript (AAM) version arising from this submission. We acknowledge the insights provided by Janusz Sylwester and Barbara Sylwester, Space Research Centre, Polish Academy of Sciences, Poland. We also thank Ewan Dickson and Hualin Xiao for providing prompt support with the \textit{STIX} data processing. Open data policy of various space-based and ground-based observatories namely \textit{Solar Orbiter}, \textit{SDO}, \textit{STEREO}, and \textit{GOES} is acknowledged. The use of the Helioviewer project, an open-source project for the visualization of solar and heliospheric data, is also acknowledged.
\end{acknowledgements}

\vspace{5mm}
\facilities{STIX (Solar Orbiter), GOES, SDO, STEREO-A}

 \software{SolarSoftWare \citep{Freeland1998, Freeland2012},
           Sparse DEM inversion technique \citep{Cheng2012, Su2018}, 
           Solar-MACH \citep{Gieseler2023} 
          }


\end{document}